\documentstyle[12pt]{ioplppt}
 
\eqnobysec

\ifx\fiverm\undefined
	\newfont\fiverm{cmr5}
\fi
\input prepictex
\input pictex
\input postpictex

\begin{document} 
\title{Ground State Structure and Low Temperature Behaviour of an
Integrable Chain with Alternating Spins}[XXZ($\frac{1}{2},1$)
Heisenberg spin chain] \author{B - D D\"orfel\footnote{email:
doerfel@qft2.physik.hu-berlin.de} and St Mei\ss ner\footnote{email:
meissner@qft2.physik.hu-berlin.de}} \address{Institut f\"ur Physik,
Humboldt-Universit\"at , Theorie der Elementarteilchen\\
Invalidenstra\ss e 110, 10115 Berlin, Germany}

\begin{abstract} 
In this paper we continue the investigation of an anisotropic integrable spin
chain, consisting of spins $s=1$ and $s=\frac{1}{2}$, started in our paper 
\cite{meissner}. The thermodynamic Bethe ansatz is analysed especially for the
case, when the signs of the two couplings $\bar{c}$ and $\tilde{c}$ differ.
For the conformally invariant model ($\bar{c}=\tilde{c}$) we have calculated
heat capacity and magnetic susceptibility at low temperature. In the isotropic
limit our analysis is carried out further and susceptibilities are calculated
near phase transition lines (at $T=0$). 
\end{abstract} 


\maketitle 
\section{Introduction}
Since the pioneering work of de Vega and Woynarovich \cite{devega} for the 
construction of models with alternating spins quite a lot interesting 
generalizations have been presented \cite{aladim1,aladim2,martins}. Otherwise, 
not so much results were obtained concerning the physical structure of the 
models, e.g. the low temperature behaviour of heat capacity and magnetic 
susceptibility. Even the structure of the ground state in the framework of the 
Bethe ansatz is not fully known for the original model.

In this paper we therefore continue our investigation of the $XXZ(\frac{1}{2}
,1)$ model with strictly alternating spins started in paper \cite{meissner},
which will be lower referred to as paper I.

In section 3 the thermodynamic Bethe ansatz (TBA) is analysed for zero 
temperature in different regions of coupling constants.

Section 4 deals with the conformally invariant model, where the low temperature
behaviour can be determined analytically.

In section 5 we derive some new results for the isotropic case 
$XXX(\frac{1}{2},1)$.

Our conclusions are contained in section 6.
\section{Definition of the model}
We send the reader to papers I and \cite{devega} for the basics of the model.
We will follow below the definitions and notations of paper I.

Our Hamiltonian of a spin chain of length $2N$ is given by
\begin{equation}\label{ham}
{\cal H}(\gamma) = \bar{c} \bar{\cal H}(\gamma) + \tilde{c} \tilde{\cal H}
(\gamma) - H S^z,
\end{equation}
with the two real coupling constants $\bar{c}$ and $\tilde{c}$.
Anisotropy parameter $\gamma$ is limited $0<\gamma<\pi/2$.

For convenience we repeat the Bethe ansatz equations (BAE) and the magnon 
energies:
\begin{equation}\label{bae}
\fl \left( \frac{\sinh(\lambda_j+i\frac{\gamma}{2})}{\sinh(\lambda_j-i
\frac{\gamma}{2})}
\frac{\sinh(\lambda_j+i\gamma)}{\sinh(\lambda_j-i\gamma)} \right)^N =
-\prod_{k=1}^{M}\frac{\sinh(\lambda_j-\lambda_k+i\gamma)}{\sinh(\lambda_j-
\lambda_k-i\gamma)},\qquad j=1\dots M,
\end{equation}
\begin{eqnarray}\label{en}
E = \bar{c} \bar{E} + \tilde{c} \tilde{E} - \left( \frac{3N}{2} - M \right) H,
\end{eqnarray}
\begin{eqnarray}\label{en1}
\bar{E} = - \sum_{j=1}^{M} \frac{2\sin\gamma}
{\cosh2\lambda_j - \cos\gamma},
\end{eqnarray}
\begin{eqnarray}\label{en2}
\tilde{E} = - \sum_{j=1}^{M} 
\frac{2\sin2\gamma}{\cosh2\lambda_j - \cos2\gamma}.
\end{eqnarray}
\section{Thermodynamic Bethe ansatz and the ground state for different signs
of the coupling constants}
In section 3 of paper I the TBA was considered for special values of 
$\gamma=\pi/\mu$, $\mu$ integer and $\mu\geq3$. We also argued, that
the ground state structure is uniform in our whole $\gamma$-region, while we
expect possible changes for the excitations at the $\gamma$-points above. We
therefore use the results of I for the possible appearance of strings in the 
ground state according to the different regions of couplings.

For completeness we quote in all cases the TBA, equations (3.19) of paper I.
We found it more convenient to use $\lambda$-space instead of Fourier 
transformation, which one easily can derive from our equations below. We then 
recall 
\begin{equation}\label{fprime}
f'(\lambda,n,\pm1) = \pm \frac{2\sin n\gamma}{\cosh 2\lambda \mp \cos n\gamma}.
\end{equation}
For shortness we drop the magnetic field in the TBA. Afterwards it can be added
without any problem. 

Now we analyse the zero temperature TBA in the various regions of signs for
$\bar{c}$ and $\tilde{c}$.

(i) $\bar{c}>0$, $\tilde{c}>0$

\begin{eqnarray}\label{tba1}
\fl
\epsilon_1^+ (\lambda) = -\bar{c} f'(\lambda,1,1) -\tilde{c}f'(\lambda,2,1)
\nonumber\\
-\left[ \delta(\lambda) + \frac{f'(\lambda,2,1)}{2\pi} \right]*\epsilon_1^-
-\left[ \frac{f'(\lambda,1,1)+f'(\lambda,3,1)}{2\pi} \right]*\epsilon_2^-,
\end{eqnarray}
\begin{eqnarray}\label{tba2}
\fl
\epsilon_2^+ (\lambda) = -\bar{c} f'(\lambda,2,1) -\tilde{c}[f'(\lambda,1,1)+
f'(\lambda,3,1)]
\nonumber\\
-\left[ \frac{f'(\lambda,1,1)+f'(\lambda,3,1)}{2\pi} \right]*\epsilon_1^-
-\left[ \delta(\lambda) + \frac{2f'(\lambda,2,1)+f'(\lambda,4,1)}{2\pi} \right]
*\epsilon_2^-.
\nonumber\\
\end{eqnarray}
The solution has been already given in \cite{devega}, where also the 
excitations have been found.

(ii) $\bar{c}>0$, $\tilde{c}<0$

We expect (1,+) and (1,-) strings.
\begin{eqnarray}\label{tba3}
\fl
\epsilon_1^+ (\lambda) = -\bar{c} f'(\lambda,1,1) -\tilde{c}f'(\lambda,2,1)
\nonumber\\
-\left[ \delta(\lambda) + \frac{f'(\lambda,2,1)}{2\pi} \right]*\epsilon_1^-
+\left[ \frac{f'(\lambda,2,-1)}{2\pi} \right]*\epsilon_{-1}^-,
\end{eqnarray}
\begin{eqnarray}\label{tba4}
\fl
\epsilon_{-1}^+ (\lambda)= -\bar{c} f'(\lambda,1,-1) -\tilde{c}f'(\lambda,2,-1)
\nonumber\\
-\left[ \frac{f'(\lambda,2,-1)}{2\pi} \right]*\epsilon_1^-
-\left[ \delta(\lambda) - \frac{f'(\lambda,2,1)}{2\pi} \right]*\epsilon_{-1}^-.
\end{eqnarray}
At the first hand, one might expect that the solution is given when both 
strings are distributed with infinite Fermi radius. We have determined this 
state and calculated its energy. But it is not the ground state. The same 
applies to the state with only (1,+) strings. That can be seen already 
superficially after obtaining $S_z\not=0$ for it.

The situation changes when only (1,-) strings are considered. This is due to 
the fact, that the two last terms in equation (\ref{tba3}) are definitely 
non-negative while this is not the case in equation (\ref{tba4}), where the 
term after the $\delta$-function spoils the argument.

Equation (\ref{tba4}) for $\epsilon_1^-(\lambda)\equiv0$ has been already 
solved in I.
\begin{eqnarray}
\fl
\epsilon_{-1}^-(\lambda) = \frac{\pi\bar{c}}{\pi-\gamma}
\frac{1}{\cosh(\pi\lambda/(\pi-\gamma))}
\nonumber\\
+ \frac{4\pi\tilde{c}}{\pi-\gamma}\frac{\cos(\pi\gamma/2(\pi-\gamma))
\cosh(\pi\lambda/(\pi-\gamma))}
{\cosh(2\pi\lambda/(\pi-\gamma))+\cos(\pi\gamma/(\pi-\gamma))}.
\end{eqnarray}
Introducing the function $g(\lambda,\alpha)$
\begin{equation}\label{g}
g(\lambda,\alpha)=\frac{4\pi}{\pi-\gamma}\frac{\cos(\pi\alpha/2(\pi-\gamma))
\cosh(\pi\lambda/(\pi-\gamma))}
{\cosh(2\pi\lambda/(\pi-\gamma))+\cos(\pi\alpha/(\pi-\gamma))}
\end{equation}
the solution of equation (\ref{tba3}) can be written as
\begin{equation}
\epsilon_1^+(\lambda) = -\bar{c} g(\lambda,\pi/2-\gamma) - \tilde{c} 
g(\lambda,\pi/2-3\gamma/2).
\end{equation}
For consistency it is necessary to have
\begin{equation}\label{4quadcond}
\epsilon_{-1}^-(\lambda) \leq 0 \quad \mbox{and} \quad \epsilon_1^+(\lambda) 
\geq 0.
\end{equation}
Both conditions specify the region of $\bar{c}$ and $\tilde{c}$ where our
solution is valid.

We start with $\epsilon_1(\lambda)$.
\begin{equation}\label{ccond1}
\fl
\epsilon_1(0) = - \frac{2\pi}{\pi-\gamma} \left[ \frac{\bar{c}}
{\cos(\pi(\pi-2\gamma)/2(\pi-\gamma))} + \frac{\tilde{c}}
{\cos(\pi(\pi-3\gamma)/2(\pi-\gamma))} \right] \geq 0.
\end{equation}
Considering the asymptotics for $\lambda\to\infty$ one has
\begin{equation}\label{ccond2}
\fl
- \frac{2\pi}{\pi-\gamma} \left[ \bar{c}\cos\frac{\pi(\pi-2\gamma)}
{2(\pi-\gamma)} 
+ \tilde{c}\cos\frac{\pi(\pi-3\gamma)}{2(\pi-\gamma)} \right] \geq 0.
\end{equation}
We now assume that the two necessary conditions (\ref{ccond1}) and 
(\ref{ccond2}) are also sufficient to fulfill the second part of 
(\ref{4quadcond}).

The smaller one of the ratios of the two cosine-functions is then the upper 
limit of $\bar{c}/|\tilde{c}|$. Hence after elementary recasting
\begin{eqnarray}\label{finc1}
\frac{\bar{c}}{|\tilde{c}|} \leq \frac{1}{2\cos(\pi\gamma/2(\pi-\gamma))},
\quad &0 \leq \gamma \leq \frac{2\pi}{5},
\nonumber\\
\frac{\bar{c}}{|\tilde{c}|} \leq 2\cos\frac{\pi\gamma}{2(\pi-\gamma)},
& \frac{2\pi}{5} \leq \gamma < \frac{\pi}{2}.
\end{eqnarray}
We treat $\epsilon_{-1}^-(\lambda)$ in the same way obtaining
\begin{equation}\label{finc2}
\frac{\bar{c}}{|\tilde{c}|} \leq 2\cos\frac{\pi\gamma}{2(\pi-\gamma)}.
\end{equation}
Now it is not difficult to show that condition (\ref{finc2}) is fulfilled, when
(\ref{finc1}) holds.

Therefore, our solution, a sea of (1,-) strings with infinite Fermi zone, is 
the ground state configuration as long as the inequalities (\ref{finc1}) hold.
In the $(\tilde{c},\bar{c})$-plane this is an open triangle formed by the 
negative $\tilde{c}$-axis and the straight line given by relation (\ref{finc1})
when the equality holds (s. figure 1). For $\gamma\to 0$ (isotropic case, s. 
section 5) this is $\bar{c}/|\tilde{c}|=\frac{1}{2}$. For increasing $\gamma$
the region first enlarges until $\gamma=2\pi/5$ and then shrinks and 
approaches the $\tilde{c}$-axis when $\gamma\to\pi/2$.

Above that line we expect still (1,-) strings but together with (1,+) strings.
So moving counter-clockwise from the positive $\bar{c}$-axis towards that line 
the Fermi radius of the strings with positive parity shrinks from infinity to
zero, while the radius for the strings with negative parity is infinite, as can
easily be seen from equations I (3.17), which implies in the case $H=0$,that 
its energy function does not change sign and is therefore strictly non-positive
in the limit $T\to0$.

It is remarkable, that a finite Fermi zone occurs without the presence of a 
magnetic field. Apparently the second coupling plays the role of an external 
field.

(iii) $\bar{c}<0$, $\tilde{c}>0$

We expect (2,+) and (1,-) strings.
\begin{eqnarray}\label{tba5}
\fl
\epsilon_2^+ (\lambda) = -\bar{c} f'(\lambda,2,1) -\tilde{c}[f'(\lambda,1,1) 
+ f'(\lambda,3,1)]
\nonumber\\
-\left[ \delta(\lambda) +\frac{2f'(\lambda,2,1)+f'(\lambda,4,1)}{2\pi} \right]*
\epsilon_2^-
\nonumber\\
+\left[ \frac{f'(\lambda,1,-1)+f'(\lambda,3,-1)}{2\pi} \right]*\epsilon_{-1}^-,
\end{eqnarray}
\begin{eqnarray}\label{tba6}
\fl
\epsilon_{-1}^+ (\lambda)= -\bar{c} f'(\lambda,1,-1) -\tilde{c}f'(\lambda,2,-1)
\nonumber\\
-\left[ \frac{f'(\lambda,1,-1)+f'(\lambda,3,-1)}{2\pi} \right]*\epsilon_2^-
-\left[ \delta(\lambda) - \frac{f'(\lambda,2,-1)}{2\pi} \right]*
\epsilon_{-1}^-.
\end{eqnarray}
We have found that qualitatively the same arguments apply as in the case (ii)
before. Thus, we consider first only (1,-) strings with infinite Fermi radius.
Now it is necessary to assure $\epsilon_2^+(\lambda)\geq0$ in addition to the 
first condition of (\ref{4quadcond}). Instead of condition (\ref{finc2}) it 
gives now
\begin{equation}\label{ccond3}
\frac{|\bar{c}|}{\tilde{c}}\geq \frac{2}{\cos(\pi\gamma/2(\pi-\gamma))},
\end{equation}
which guarantees $\epsilon^-_{-1}(\lambda)\leq0$.
When calculating $\epsilon^+_2$ one has to be careful when the Fourier 
transformation of $f'(\lambda,3,-1)$ is to be taken. It vanishes for 
$\gamma=\pi/3$ and changes the sign after that point had been passed. Finally 
one obtains
\begin{eqnarray}
\fl
\epsilon^+_2(\lambda)= - \bar{c} g(\lambda,\pi/2-3\gamma/2) - \tilde{c} 
g(\lambda,\pi/2-\gamma) - \tilde{c} g(\lambda,\pi/2-2\gamma),
&0 < \gamma < \pi/3,
\nonumber\\
\fl
\epsilon^+_2(\lambda) \equiv 0, &\pi/3 < \gamma < \pi/2.
\end{eqnarray}
There is no contradiction with formulae I (3.24), which gives two different 
values for $\gamma<\pi/3$ and $\gamma=\pi/3$, while larger $\gamma$-values were
not considered there.

Let us first consider $0<\gamma<\pi/3$. Then from equation (\ref{ccond3}) we 
have the two conditions
\begin{eqnarray}\label{ccond4}
\fl
-\frac{2\pi}{\pi-\gamma}\left(\frac{\bar{c}}{\cos(\pi(\pi-3\gamma)
/2(\pi-\gamma))} \right. 
\nonumber\\
\left.
+ \frac{\tilde{c}}{\cos(\pi(\pi-2\gamma)/2(\pi-\gamma))}
+ \frac{\tilde{c}}{\cos(\pi(\pi-4\gamma)/2(\pi-\gamma))}\right)\geq0,
\nonumber\\
\fl
-\frac{2\pi}{\pi-\gamma}\left(\bar{c}\cos\frac{\pi(\pi-3\gamma)}
{2(\pi-\gamma)} \right. 
\nonumber\\
\left.
+ \tilde{c}\cos\frac{\pi(\pi-2\gamma)}{2(\pi-\gamma)}
+ \tilde{c}\cos\frac{\pi(\pi-4\gamma)}{2(\pi-\gamma)}\right)\geq0.
\end{eqnarray}
Straightforward calculation gives
\begin{eqnarray}\label{finc3}
\frac{|\bar{c}|}{\tilde{c}}\geq 2\cos \frac{\pi\gamma}{2(\pi-\gamma)} \quad 
\mbox{and}
\nonumber\\
\frac{|\bar{c}|}{\tilde{c}}\geq \frac{8\cos^3(\pi\gamma/2(\pi-\gamma))}
{4\cos^2(\pi\gamma/2(\pi-\gamma))-1}.
\end{eqnarray}
The upper term of the RHS is always smaller than the RHS of (\ref{ccond3}). 
Hence we have to find the maximum of the two RHS of formula (\ref{finc3}) and
(\ref{ccond3}). In our $\gamma$-region the second inequality of (\ref{finc3})
is the most restrictive one. Putting things together we find for the region 
with (1,-) strings only
\begin{eqnarray}\label{finc4}
\frac{|\bar{c}|}{\tilde{c}}\geq \frac{8\cos^3(\pi\gamma/2(\pi-\gamma))}
{4\cos^2(\pi\gamma/2(\pi-\gamma))-1}, \quad &0<\gamma\leq\frac{\pi}{3},
\nonumber\\
\frac{|\bar{c}|}{\tilde{c}}\geq \frac{2}{\cos \pi\gamma/2(\pi-\gamma)},
\quad &\frac{\pi}{3}\leq\gamma<\frac{\pi}{2}. 
\end{eqnarray}
In the $(\tilde{c},\bar{c})$-plane that is an open triangle formed by the 
negative $\bar{c}$-axis and the straight line given by relation (\ref{finc4})
when the equality holds (s. figure 1). For $\gamma\to0$ (isotropic case, s. 
section 5) this is $|\bar{c}|/\tilde{c}=\frac{8}{3}$. For rising $\gamma$ the
region shrinks and approaches the $\tilde{c}$-axis when $\gamma\to\pi/2$.

Above that region we expect (1,-) strings together with (2,+) strings the 
latter with finite Fermi radius. The picture resembles region (ii) treated 
before.

(iv) $\bar{c}\leq0$, $\tilde{c}\leq0$

Here the vacuum is formed by (1,-) strings only.
\begin{eqnarray}\label{tba7}
\fl
\epsilon_{-1}^+(\lambda)= -\bar{c} f'(\lambda,1,-1) -\tilde{c} f'(\lambda,2,-1)
-\left[ \delta(\lambda) + \frac{f'(\lambda,2,-1)}{2\pi} \right]*
\epsilon_{-1}^-.
\end{eqnarray}
This region was studied in I where also the excitations have been found.

(v) $\bar{c}=0$, $\tilde{c}>0$

(vi) $\bar{c}>0$, $\tilde{c}=0$

We add nothing new to both cases considered earlier in \cite{devega} and 
\cite{devega1}.

Now we can summarize our results about the the ground state structure for 
different values of coupling constants. There are four regions and two singular
lines (v) and (vi). In the two regions with equal signs (which contain the line
$\bar{c}=\tilde{c}$) the ground state is independent of the values of $\bar{c}$
and $\tilde{c}$. Here also the Fermi radii are infinite. There is no mass gap 
in the excitation spectrum.

In the two other regions infinite and finite Fermi radii occur and the concrete
structure of the ground state depends on the ratio $\bar{c}/\tilde{c}$. 
Nevertheless, we expect them to be gapless, too.

The picture is not fully symmetric, because region (i) is separated from all 
others by highly degenerate ground state on both lines. This is connected with 
the fact, that one sort of strings has to disappaear at once.

Finally, the model shows an antiferromagnetic behaviour everywhere (for 
vanishing magnetic field) as long as $\gamma>0$. The isotropic case is 
considered in section 5.

\begin{figure}
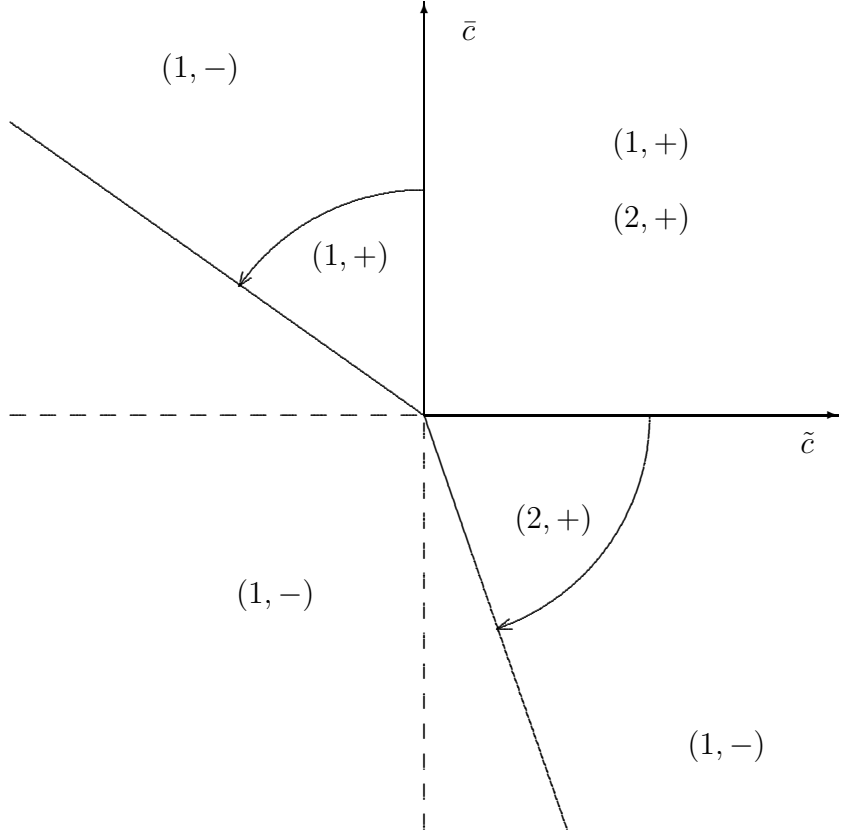

\input figure1.tex
\caption{The phase structure of the $XXZ(\frac{1}{2},1)$ model for 
$\gamma=\pi/3$. The ground state strings are indicated for the four different
sectores. The arrows symbolize the decreasing Fermi radii of the corresponding
strings. Broken lines are coordinate axes. Where axes are drawn solid they 
coincide with sector borders.}
\end{figure}

\section{Calculation of the low temperature behaviour in the case 
$\bar{c}=\tilde{c}$}
In this section we calculate the low temperature heat capacity and magnetic 
susceptibility for vanishing magnetic field in the case $\bar{c}=\tilde{c}$.

We therefore go back to equations (3.10)-(3.13) of paper I where $T$ is 
considered to be small but finite. Instead of paper \cite{devega1} where the 
free energy was calculated, we use a method due to Wiegmann \cite{wieg},
which for our purpose was used by Babujian and Tsvelick \cite{tsve} to obtain
the results for the $XXZ(S)$ model from entropy and polarization. To explore 
this method it is necessary to ensure $\gamma<\pi/3$, $\gamma=\pi/\mu$; $\mu$
integer.

We will present in some detail the case $c=\bar{c}=\tilde{c}<0$ while for 
$c>0$ we mention only the necessary changes and the final results.

For $c<0$ we have
\begin{eqnarray}
\epsilon_j\geq0, \quad j=1 \dots \mu-1,
\nonumber\\
\epsilon_{\mu}\leq0,\nonumber
\end{eqnarray}
and
\begin{eqnarray}
\rho_j \to0 \quad \mbox{for} \quad T\to0 \quad \mbox{if} \quad j=1 \dots \mu-1,
\nonumber\\
\tilde{\rho}_{\mu} \to 0 \quad \mbox{for} \quad T\to0.\nonumber
\end{eqnarray}
Our aim is now to recast BAE and TBAE in a form where energies and densities 
are given through their zero temperature limits $\epsilon_{\mu}^{(0)}$ and 
$\epsilon_j^{(0)}$ and values vanishing for $T\to0$, i. e. the energy functions
$\ln(1+\exp(-\epsilon_j/T)),\quad j=1\dots\mu-1$ and 
$\ln(1+\exp(\epsilon_{\mu}/T))$.

The main step is the multiplication by $A_{\mu\mu}^{-1}$ which after some 
algebra leads to the systems
\begin{eqnarray}\label{tbae_recast}
\fl
\epsilon_{\mu} = \frac{H\mu}{2} \epsilon_{\mu}^{(0)} - \sum_{k=1}^{\mu-1}
Q_k * (-1)^{r(k)} T& \ln\left(1+\exp\left( -\frac{\epsilon_k}{T} \right)\right)
\nonumber\\
&- K * (-1)^{r(\mu)} T \ln\left(1+\exp\left( \frac{\epsilon_{\mu}}{T} \right)
\right),
\nonumber
\end{eqnarray}
\begin{eqnarray}
\fl
\epsilon_j = \delta_{j\mu-1}\frac{H\mu}{2} \epsilon_j^{(0)} 
+ \sum_{k=1}^{\mu-1}
B_{jk} * (-1)^{r(k)} T&\ln\left(1+\exp\left(-\frac{\epsilon_k}{T}\right)\right)
\nonumber\\
&- K * (-1)^{r(\mu)} T \ln\left(1+\exp\left( \frac{\epsilon_{\mu}}{T} \right)
\right)
\end{eqnarray}
and
\begin{eqnarray}\label{bae_recast}
-(-1)^{r(\mu)} (\tilde{\rho}_{\mu} + \rho_{\mu}) = 
\frac{1}{2\pi c} \epsilon_{\mu}^{(0)} - \sum_{k=1}^{\mu-1} Q_k * \rho_k 
+ K * \tilde{\rho}_{\mu},
\nonumber\\
-(-1)^{r(j)} (\tilde{\rho}_j + \rho_j) = \frac{1}{2\pi c} \epsilon_j^{(0)} 
+ \sum_{k=1}^{\mu-1} B_{jk} * \rho_k + Q_j * \rho_{\mu},
\end{eqnarray}
where we have introduced 
\begin{eqnarray}
K(\lambda) = - T_{\mu\mu} * A_{\mu\mu}^{-1} (\lambda)
\nonumber\\
Q_k (\lambda) = - T_{\mu k} * A_{\mu\mu}^{-1} (\lambda)
\nonumber\\
B_{jk} (\lambda) = (T_{j k} + T_{j \mu} * A_{\mu\mu}^{-1}* T_{\mu k})(\lambda).
\end{eqnarray}
Now we want to perform a shift of the $\lambda$-variable in the functions 
$\epsilon_j(\lambda)$ in the following way:
\begin{equation}\label{shift}
\varphi_j (\lambda) = \frac{1}{T} \epsilon_j\left(\lambda + 
a\ln \frac{T}{2\pi|c|}\right)
\end{equation} 
where the constant $a$ will be determined yet. We choose it in a way that 
$\varphi_j(\lambda)$ for $\lambda\to\infty$ has a finite limit if $T\to0$.
From I (3.24) we can see, that for all $\epsilon_j^{(0)}(\lambda) \sim 
\exp(\pi\lambda/(\pi-\gamma))$ if $\lambda\to\infty$. Therefore 
$a=-(\pi-\gamma)/\gamma$.

After that shift the system (\ref{tbae_recast}) is rewritten as
\begin{eqnarray}\label{phi_tbae}
\fl
\varphi_{\mu} = \frac{H\mu}{2T} - \sum_{k=1}^{\mu-1}
Q_k * (-1)^{r(k)} \ln(1+\exp\left( -\varphi_k \right))
- K * (-1)^{r(\mu)} \ln(1+\exp\left( \varphi_{\mu} \right))
\nonumber\\
+ \frac{1}{\pi-\gamma}\exp\left(-\frac{\pi\lambda}{\pi-\gamma}\right)
\left[1+2\cos\frac{\pi}{2(\mu-1)}\right],
\nonumber\\
\fl
\varphi_j = \delta_{j\mu-1}\frac{H\mu}{2T}
+ \sum_{k=1}^{\mu-1}
B_{jk} * (-1)^{r(k)} \ln(1+\exp\left( -\varphi_k \right))
- K * (-1)^{r(\mu)} \ln(1+\exp\left( \varphi_{\mu} \right))
\nonumber\\
+ \frac{2}{\pi-\gamma}\exp\left(-\frac{\pi\lambda}{\pi-\gamma}\right)\times
\nonumber\\
\times
\left[\cos\frac{(\mu-j-1)\pi}{2(\mu-1)} + \cos\frac{(\mu-j)\pi}{2(\mu-1)}
+ \cos\frac{(\mu-j-1)\pi}{2(\mu-2)}\right].
\end{eqnarray}
The system (\ref{bae_recast}) is treated analogously. 

After differentiating equation (\ref{phi_tbae}) one obtains the important 
relations
\begin{eqnarray}\label{relations}
\rho_j\left(\lambda - \frac{\mu-1}{\mu} \ln \frac{T}{2\pi|c|}\right) =
(-1)^{r(j)}\frac{\pi-\gamma}{2\pi^2|c|} T \frac{\partial}{\partial\lambda}
\ln \left( 1+\exp(-\varphi_j) \right),
\nonumber\\
\tilde{\rho}_j\left(\lambda - \frac{\mu-1}{\mu} \ln \frac{T}{2\pi|c|}\right) =
-(-1)^{r(j)}\frac{\pi-\gamma}{2\pi^2|c|} T \frac{\partial}{\partial\lambda}
\ln \left( 1+\exp(\varphi_j) \right)
\end{eqnarray}
for $j=1\dots\mu$.

These relations are necessary to make the appropriate substitutions of 
variables in the integrals for $S$ and $S_z$. No such relations are expected as
soon as $\bar{c}\not=\tilde{c}$.

The starting point for the heat capacity calculation is the expression for the
entropy
\begin{equation}\label{entropy}
\frac{S}{N} = \sum_{j=1}^{\mu} \int_{-\infty}^{\infty}d\lambda
\left[ \rho_j\ln\left( 1+\frac{\tilde{\rho}_j}{\rho_j} \right) +
\tilde{\rho}_j \ln\left( 1+\frac{\rho_j}{\tilde{\rho}_j} \right) \right].
\end{equation}
Using symmetry and $\tilde{\rho}_j/\rho_j=e^{\epsilon_j/T}$ we have
\begin{equation}\label{entropy1}
\frac{S}{N} = 2 \sum_{j=1}^{\mu} \int_0^{\infty}d\lambda
\left[ \rho_j\ln\left( 1+ e^{\epsilon_j/T} \right) +
\tilde{\rho}_j\ln\left( 1+  e^{-\epsilon_j/T}\right) \right].
\end{equation}
For $T\to0$ the main contribution to the integral comes from $\lambda\gg1$.

After performing the shift (\ref{shift}) and using relations (\ref{relations})
the entropy becomes
\begin{eqnarray}\label{entropy2}
\fl
\frac{S}{N} = \frac{\pi-\gamma}{\pi^2|c|} \sum_{j=1}^{\mu} (-1)^{r(j)}
\int_{\frac{\mu-1}{\mu}\ln\left(\frac{T}{2\pi|c|}\right)}^{\infty}d\lambda
\left[ \frac{\partial}{\partial\lambda}\ln\left( 1+ e^{-\varphi_j} \right)
\ln\left( 1+ e^{\varphi_j} \right) \right.
\nonumber\\
\left.
+\frac{\partial}{\partial\lambda}\ln\left( 1+  e^{\varphi_j} \right) 
\ln\left( 1+ e^{-\varphi_j} \right) \right].
\end{eqnarray}
We are interested only in the leading order for vanishing temperature. 
Therefore we can substitute the lower limit by $-\infty$ (both integrals 
converge).

We will see in a moment that the remaining integral is even independent of $T$.

Now it is straightforward to change the variable in the way
\begin{equation}\label{varchange}
x=\frac{1}{1+e^{\varphi_j}}\equiv f(\varphi_j)
\end{equation}
for every integral in the sum (Note the change in the definition of the 
function $f$ in equation (3.18) of paper I).

In final form
\begin{eqnarray}\label{entropy3}
\frac{S}{N} = -\frac{\pi-\gamma}{\pi^2|c|} \sum_{j=1}^{\mu} (-1)^{r(j)}
\int_{f(\varphi_j^-)}^{f(\varphi_j^+)}dx
\left[ \frac{\ln x}{1-x} + \frac{\ln(1-x)}{x} \right]
\end{eqnarray}
with $\varphi_j^{\pm}=\varphi_j(\pm\infty)$.

The integral is given by the function $\gamma(a,b)$ already introduced in 
\cite{tsve} from where also the necessary special values have been taken.
\begin{equation}\label{gamma}
\gamma(a,b)=\int_a^b dx \left[ \frac{\ln x}{1-x} + \frac{1}{x}\ln(1-x) \right].
\end{equation}
Following the standard procedure \cite{wieg} it is more convenient to use 
another form of the TBAE to determine $\varphi_j^{\pm}$. They differ from those
of the $XXZ(S)$ model \cite{tsve} in the terms with the coupling constants 
only. It is the system (3.17) from paper I.

After the shift it takes the form
\begin{eqnarray}\label{tbae_invert}
\varphi_1(\lambda) = - s*\ln f(\varphi_2)(\lambda) + |c|\exp\left(-\frac
{\pi\lambda}{\pi-\gamma}\right)
\nonumber\\
\varphi_j(\lambda) = - s*\ln [f(\varphi_{j+1})f(\varphi_{j-1})](\lambda) 
+ |c|\exp\left(-\frac{\pi\lambda}{\pi-\gamma}\right)\delta_{j2}
\nonumber\\
\varphi_{\mu-1}(\lambda) = \frac{H\mu}{2T} - s*\ln f(\varphi_{\mu-2})(\lambda)
\nonumber\\
\varphi_{\mu}(\lambda) = \frac{H\mu}{2T} + s*\ln f(\varphi_{\mu-2})(\lambda).
\end{eqnarray} 
For $\lambda\to-\infty$ the inhomogeneous terms generate a solution of the form
\begin{eqnarray}\label{phi-}
\varphi_j^- = + \infty, \quad j=1\dots\mu-1,
\nonumber\\
\varphi_{\mu}^- = - \infty,
\end{eqnarray}
which implies
\begin{eqnarray}\label{f-}
f(\varphi_j^-) = 0, \quad j=1\dots\mu-1,
\nonumber\\
f(\varphi_{\mu}^-) = 1.
\end{eqnarray}
For $\lambda\to\infty$ the free terms can be neglected, and thus the solution 
is given in \cite{tsve}
\begin{eqnarray}\label{phi+}
f(\varphi_j^+) = \left[ \frac{\sinh(H/2T)}{\sinh(H(j+1)/2T)} \right]^2,
\nonumber\\
\varphi_{\mu-1}^+ = \frac{H\mu}{2T} + 
\ln \left[ \frac{\sinh(H(\mu-1)/2T)}{\sinh(H/2T)} \right],
\nonumber\\
\varphi_{\mu}^+ = \frac{H\mu}{2T} -
\ln \left[ \frac{\sinh(H(\mu-1)/2T)}{\sinh(H/2T)} \right].
\end{eqnarray} 
For $H\to0$ then
\begin{eqnarray}\label{f+}
f(\varphi_j^+) = \frac{1}{(j+1)^2},\quad
f(\varphi_{\mu-1}^+) = \frac{1}{\mu},\quad
f(\varphi_{\mu}^+) = 1 - \frac{1}{\mu}.
\end{eqnarray}
We mention that the above solution does not depend on the sign of the coupling 
constant. The consequences of that fact will be considered below.

Now we can calculate relation (\ref{entropy3}):
\begin{eqnarray}\label{entropy4}
\frac{S}{N} &=& \frac{(\pi-\gamma)T}{\pi^2|c|}\left\{ \sum_{j=1}^{\mu-2}
\gamma\left(\frac{1}{(j+1)^2},0\right) + \gamma\left(\frac{1}{\mu},0\right) - 
\gamma\left(1-\frac{1}{\mu},1\right) \right\}
\nonumber\\
&=& \frac{(\pi-\gamma)T}{\pi^2|c|}\left\{ \sum_{j=1}^{\mu-2}
\gamma\left(\frac{1}{(j+1)^2},0\right) + 2\gamma\left(\frac{1}{\mu},0\right) 
\right\}
\nonumber\\
&=& \frac{(\pi-\gamma)T}{\pi^2|c|}\left\{ \frac{1}{3}\pi^2 \right\}
\nonumber\\
&=& \frac{(\pi-\gamma)T}{3|c|}.
\end{eqnarray}
Finally, for the heat capacity per site (of a chain with $2N$ sites)
\begin{equation}\label{heat}
C =  \frac{\pi-\gamma}{6|c|}T.
\end{equation}
The polarization is obtained in the same way as in \cite{tsve} starting with 
the basic formula
\begin{eqnarray}\label{spin}
\frac{S^z}{N} = \frac{\mu}{2} \left\{ \int_{-\infty}^{\infty} d\lambda
\tilde{\rho}_{\mu-1}(\lambda) - \int_{-\infty}^{\infty} d\lambda
\rho_{\mu}(\lambda)\right\}
\end{eqnarray}
which for our model takes the same form. Using symmetry and the shift gives
\begin{eqnarray}\label{spin1}
\fl
\frac{S^z}{N} &= 2\frac{\mu}{2} \left\{ \int_0^{\infty} d\lambda
\tilde{\rho}_{\mu-1}(\lambda) - \int_0^{\infty} d\lambda
\rho_{\mu}(\lambda)\right\}
\nonumber\\
\fl
&= \mu \left\{ -\int_{-\infty}^{\infty} \frac{(\pi-\gamma)T}{2\pi^2|c|}
\frac{\partial}{\partial\lambda}\ln\left( 1+ e^{\varphi_{\mu-1}} \right)
d\lambda
+ -\int_{-\infty}^{\infty} \frac{(\pi-\gamma)T}{2\pi^2|c|}
\frac{\partial}{\partial\lambda}\ln\left( 1+ e^{-\varphi_{\mu}} \right)
d\lambda \right\}
\nonumber\\
\fl
&= \frac{\mu(\pi-\gamma)T}{2\pi^2|c|} \left\{ \ln \left[ \frac{1+
\exp(-\varphi_{\mu}^+)}{1+\exp(\varphi_{\mu-1}^+)} \right] 
- \ln \left[ \frac{1+\exp(-\varphi_{\mu}^-)}{1+\exp(\varphi_{\mu-1}^-)} 
\right] \right\}. 
\end{eqnarray}
For the first term we use relations (\ref{phi+}) with $H\gg T$.
\begin{eqnarray}\label{phi+_HggT}
\varphi_{\mu-1}^+ = \frac{H\mu}{2T} + 
\ln \left[ \frac{\exp(H(\mu-1)/2T)}{\exp(H/2T)} \right]
=\frac{H(\mu-1)}{T},
\nonumber\\
\varphi_{\mu}^+ = \frac{H\mu}{2T} -
\ln \left[ \frac{\exp(H(\mu-1)/2T)}{\exp(H/2T)} \right]
=\frac{H}{T}.
\end{eqnarray}
and hence
\begin{eqnarray}\label{f+_HggT}
1+\exp(\varphi_{\mu-1}^+) = \exp\left(\frac{H(\mu-1)}{T}\right),
\nonumber\\
1+\exp(-\varphi_{\mu}^+) = 1.
\end{eqnarray}
In the second term term (\ref{phi-}) must be completed by corrections 
containing the leading term in $H$.
\begin{eqnarray}\label{phi+corr}
\ln \left[ \frac{1+\exp(-\varphi_{\mu}^-)}{1+\exp(\varphi_{\mu-1}^-)}\right]
\cong - \varphi_{\mu}^- - \varphi_{\mu-1}^-
= - \frac{H\mu}{T}
\end{eqnarray}
Putting things together
\begin{equation}\label{spin2}
\frac{S^z}{N} = \mu \frac{(\pi-\gamma)H}{2\pi^2|c|}
\end{equation}
and finally for the susceptibility
\begin{equation}\label{susz}
\chi = \frac{\mu-1}{4\pi|c|} = \frac{\pi-\gamma}{4\pi\gamma|c|}.
\end{equation}
Now we perform the calculation for $\bar{c}=\tilde{c}=c>0$. We will list only 
the necessary changes in the calculation before, induced by the sign of the
coupling constant. For $c>0$ we have
\begin{eqnarray}
\epsilon_1,\epsilon_2 \leq 0,
\nonumber\\
\epsilon_j \geq 0, \quad j=3\dots\mu.\nonumber
\end{eqnarray}
The shift in equation (\ref{shift}) must now be taken as 
$a=-\pi/\gamma$ according to the asymptotics of the $\epsilon_j^{(0)}$.
Instead of equation (\ref{relations}) we have then
\begin{eqnarray}\label{relations1}
\rho_j\left(\lambda - \mu \ln \frac{T}{2\pi c}\right) =
(-1)^{r(j)}\frac{\gamma}{2\pi^2 c} T \frac{\partial}{\partial\lambda}
\ln \left( 1+\exp(-\varphi_j) \right),
\nonumber\\
\tilde{\rho}_j\left(\lambda - \mu \ln \frac{T}{2\pi c}\right) =
-(-1)^{r(j)}\frac{\gamma}{2\pi^2 c} T \frac{\partial}{\partial\lambda}
\ln \left( 1+\exp(\varphi_j) \right).
\end{eqnarray}
Note the change of the overall sign.

Consequently, (\ref{entropy3}) is modified
\begin{eqnarray}\label{entropy5}
\frac{S}{N} = -\frac{\gamma}{\pi^2 c} \sum_{j=1}^{\mu} (-1)^{r(j)}
\int_{f(\varphi_j^-)}^{f(\varphi_j^+)}dx
\left[ \frac{\ln x}{1-x} + \frac{\ln(1-x)}{x} \right].
\end{eqnarray}
The change in the system (\ref{tbae_invert}) is obviously the replacement of 
$|c|$ by $-c$. As already mentioned above, there are no changes in the 
solutions for $\lambda\to\infty$. For $\lambda\to -\infty$ the solution is 
contained in \cite{tsve}:
\begin{eqnarray}\label{phi-_1}
f(\varphi_1^-) = f(\varphi_2^-) = 1,
\nonumber\\
f(\varphi_j^-) = \left[ \frac{\sinh((H\mu/2T)1/(\mu-2))}{\sinh((H\mu/2T)(j-1)
/(\mu-2))} \right]^2,\quad j=3\dots\mu-2
\nonumber\\
\varphi_{\mu-1}^- = \frac{H\mu}{2T} + 
\ln \left[ \frac{\sinh((H\mu/2T)(\mu-3)/(\mu-2))}{\sinh(H\mu/2T)1/(\mu-2)} 
\right],
\nonumber\\
\varphi_{\mu}^- = \frac{H\mu}{2T} -
\ln \left[ \frac{\sinh((H\mu/2T)(\mu-3)/(\mu-2))}{\sinh(H\mu/2T)1/(\mu-2)} 
\right].
\end{eqnarray}
For $H\to0$ this implies
\begin{eqnarray}\label{f-_1}
f(\varphi_j^-) = \frac{1}{(j-1)^2},\quad j=3\dots\mu-2,
\nonumber\\
f(\varphi_{\mu-1}^-) = \frac{1}{\mu-2},\quad
f(\varphi_{\mu}^-) = 1 - \frac{1}{\mu-2}.
\end{eqnarray}
Now we are ready to find the sum in equation (\ref{entropy4})
\begin{eqnarray}\label{gammasum}
\fl
\sum_{j=1}^{\mu}(-1)^{r(j)} \gamma \left(\varphi_j^-,\varphi_j^+\right)
= \gamma\left(1,\frac{1}{4}\right) + \gamma\left(1,\frac{1}{9}\right) 
\nonumber\\
+\sum_{j=3}^{\mu-2}\gamma \left(\frac{1}{(j-1)^2},\frac{1}{(j+1)^2}\right) +
\gamma\left(\frac{1}{\mu-2},\frac{1}{\mu}\right) -
\gamma\left(1-\frac{1}{\mu-2},1-\frac{1}{\mu}\right)
\nonumber\\
= \gamma\left(1,\frac{1}{4}\right) + \gamma\left(1,\frac{1}{9}\right) + 
\sum_{j=1}^{\mu-4}\gamma \left(\frac{1}{(j+1)^2},0\right)
\nonumber\\
- \sum_{j=1}^{\mu-2}\gamma \left(0,\frac{1}{(j+1)^2}\right) 
+ \gamma\left(0,\frac{1}{4}\right) +  \gamma\left(0,\frac{1}{9}\right) + 
2 \gamma\left(\frac{1}{\mu-2},\frac{1}{\mu}\right)
\nonumber\\
= 2\gamma\left(1,0\right)+\frac{\pi^2}{3}-2\gamma\left(\frac{1}{\mu-2},0\right)
+ 2 \gamma\left(0,\frac{1}{\mu}\right) - \frac{\pi^2}{3} + 
2 \gamma\left(\frac{1}{\mu-2},\frac{1}{\mu}\right)
\nonumber\\
= 2 \gamma\left(1,0\right)
\nonumber\\
= \frac{2\pi^2}{3}.
\end{eqnarray}
Finally,
\begin{equation}\label{heat1}
C = \frac{\gamma T}{3 c}.
\end{equation}
Polarization (\ref{spin1}) is modified
\begin{eqnarray}\label{spin3}
\frac{S^z}{N} =  \frac{T}{2\pi c } \left\{ \ln \left[ \frac{1+
\exp(\varphi_{\mu-1}^+)}{1+\exp(-\varphi_{\mu}^+)} \right] 
- \ln \left[ \frac{1+\exp(\varphi_{\mu-1}^-)}{1+\exp(-\varphi_{\mu}^-)} 
\right] \right\}. 
\end{eqnarray}
Relations (\ref{f+_HggT}) are still valid. From equation (\ref{phi-_1}) we 
obtain
\begin{eqnarray}\label{phi-_1_HggT}
\varphi_{\mu-1}^- = \frac{H\mu}{T} \frac{\mu-3}{\mu-2} \quad \mbox{and}
\nonumber\\
\varphi_{\mu}^- = \frac{H\mu}{T} \frac{1}{\mu-2}.
\end{eqnarray}
Therefore,
\begin{eqnarray}\label{spin4}
\frac{S^z}{N} &=& \frac{T}{2\pi c} \left\{ \frac{H}{T} \left[ \mu - 1 - \mu +
\frac{\mu}{\mu-2} \right] \right\}
\nonumber\\
&=& \frac{T}{2\pi c} \frac{H}{T} \frac{2}{\mu-2}
\end{eqnarray}
and finally
\begin{eqnarray}\label{susz1}
\chi = \frac{1}{2\pi c} \frac{1}{\mu-2} =  \frac{1}{2\pi c} 
\frac{\gamma}{\pi-2\gamma}.
\end{eqnarray}
Now we have to compare our results with those of other authors who have 
presented calculations especially for $c>0$. To avoid ambiguities we multiply
heat capacity and susceptibility by the speed of sound $v_s$, afterwards the
result becomes unique, not depending on the normalisation of the coupling 
constant.

From \cite{meissner} and \cite{devega} we can derive
\begin{eqnarray}\label{speed_of_sound}
v_s = \frac{2 c\pi}{\gamma} \quad \mbox{for} \quad c>0 \quad \mbox{and}
\nonumber\\
v_s = \frac{2|c|\pi}{\pi-\gamma} \quad \mbox{for} \quad c<0.
\end{eqnarray}
We have used our method also to obtain the values for the two homogeneous 
systems ($s=\frac{1}{2}$ and $s=1$). At least, the susceptibility for $s=1$ has
not been calculated before in the case of negative coupling. (Heat capacity was
determined in paper \cite{alcaraz}.) In all cases considered we found for the 
heat capacity the conformal result
\begin{equation}
C v_s = \frac{c_v T \pi}{3}
\end{equation}
where $c_v$ is the central charge of the Virasoro algebra, which is equal to 
one for negative coupling.

It is remarkable that formula (\ref{entropy3}) is preserved, because 
(rewritten for the entropy per site ) the factor in front of the sum is always 
equal $1/(v_s\pi)$ (apart from sign) while the sum measures the central charge
(being equal to $\pi^2 c_v/3$).

The form of equation (\ref{spin1}) for the polarization per site can be 
understood in the same way. The factor in front of the logarithms is always
$T/(2v_s\gamma)$, while for different signs and spins the logarithms also 
differ.

For positive coupling they yield the result $2S'/(\pi-2\gamma S')H/T$ where 
$S'$ is the larger of the two spins (our model has $S'=1$), which may be equal.

Therefore
\begin{equation}
v_s \chi (c>0) = \frac{S'}{\pi-2S'\gamma}
\end{equation}
which is consistent with all former results, especially with paper \cite{tsve}
for homogeneous chains and with paper \cite{aladim2} for the isotropic limit of
alternating chains with $S'>S$.

For negative coupling the logarithms always equal $H/T$ leading to
\begin{equation}
v_s \chi (c<0) = \frac{1}{2\gamma}
\end{equation}
with no dependence on the spins. This is remarkable, because we remember 
$c_v=1$ in the same case.

\section{The isotropic model with alternating spins}
In this section we present some results for the isotropic limit of the model
considered before, which we will call $XXX(\frac{1}{2},1)$. On one side, there 
are some peculiarities in the limit $\gamma\to0$ (especially for negative 
couplings). On the other side, in the sectors with different signs of couplings
it is possible to obtain several new results yet undiscovered for the 
anisotropic case.

To begin with we have to define the isotropic limit. The model has been 
considered in paper \cite{devega1}, we have only to specify the normalization
of coupling constants to fit with our section 2.

The BAE take the form
\begin{equation}\label{iso_bae}
\left( \frac{\lambda_j+\frac{i}{2}}{\lambda_j-\frac{i}{2}}
\frac{\lambda_j+i}{\lambda_j-i} \right)^N =
-\prod_{k=1}^{M}\frac{\lambda_j-\lambda_k+i}{\lambda_j-
\lambda_k-i},\qquad j=1\dots M.
\end{equation}
Instead of (\ref{en}) we define
\begin{equation}\label{iso_en}
E = c_1 E_1 + c_2 E_2 - \left( \frac{3}{2}N - M \right) H
\end{equation}
with
\begin{equation}\label{iso_en1}
E_i = - \sum_{j=1}^M a_i(\lambda_j), \quad i=1,2
\end{equation}
where
\begin{equation}\label{a_n}
a_n(\lambda) = \frac{n}{\lambda^2 + \frac{n^2}{4}}.
\end{equation}
Taking the limit of equations (\ref{en1}) and (\ref{en2}) we see that we have 
to put 
\begin{eqnarray}\label{c_ren}
c_1 = \lim_{\gamma\to0} \frac{\bar{c}}{\gamma} \quad \mbox{and}
\nonumber\\
c_2 = \lim_{\gamma\to0} \frac{\tilde{c}}{\gamma}.
\end{eqnarray}
The TBA (for zero temperature) has been given in \cite{devega1}:
\begin{eqnarray}\label{iso_tbae}
\epsilon_1(\lambda) = - 2\pi c_1 p(\lambda) + p*\epsilon_2^+(\lambda),
\nonumber\\
\epsilon_2(\lambda) = - 2\pi c_2 p(\lambda) + p*\epsilon_1^+(\lambda) +
h*\epsilon_2^+(\lambda) + \frac{H}{2}.
\end{eqnarray}
Here
\begin{eqnarray}\label{ph}
p(\lambda) = \frac{1}{2\cosh\pi\lambda},
\nonumber\\
h(\lambda) = \int_{-\infty}^{\infty} \frac{dp}{2\pi} \frac{e^{-|p|/2}}
{2\cosh (p/2)} e^{ip\lambda}.
\end{eqnarray}
As in section 3 we must distinguish between the various regions of signs when 
the system (\ref{iso_tbae}) is solved. We will follow directly the notation 
from section 3 above.

The solutions in the sectors (i) and (iv) are well known. While in (i) the 
ground state is antiferromagnetic (1- and 2-strings) and therefore the limit of
the anisotropic case, in (iv) the ground state is ferromagnetic and hence 
different from the anisotropic case. This explains why the results of section 3
for $c>0$ (with the replacement (\ref{c_ren})) lead to the isotropic values,
while they diverge for $c<0$.

Now we wish to investigate in some detail the sector (ii) with $c_1>0,c_2<0$.
Only real roots can be present in the ground state. The TBA (only one equation 
left) can be formulated in two equivalent ways, which we shall need both.
\begin{eqnarray}\label{tbae1_ii}
\epsilon_1(\lambda) = - 2\pi c_1 p(\lambda) + 2\pi |c_2| h(\lambda) +
h*\epsilon_1^+(\lambda) + \frac{H}{2},
\end{eqnarray}
\begin{eqnarray}\label{tbae2_ii}
\epsilon_1^+(\lambda) = - c_1 a_1(\lambda) + |c_2| a_2(\lambda)
- \left[ \delta(\lambda) +\frac{a_2(\lambda)}{2\pi} \right]*\epsilon_1^- + H.
\end{eqnarray}
From equation (\ref{tbae2_ii}) one easily determines the region where the 
solution is ferromagnetic. The integral term is always non-negative. Therefore
we have ferromagnetic behaviour ($\epsilon(\lambda)>0$ everywhere) if the 
remaining function of $\lambda$ on the RHS of equation (\ref{tbae2_ii}) is 
strictly positive. That is guaranteed, if it is fulfilled for $\lambda=0$. Thus
\begin{eqnarray}\label{iso_ccond}
-4 c_1 + 2 |c_2| +H > 0
\end{eqnarray}
implying
\begin{eqnarray}\label{Hcrit}
H_{crit} =  4 c_1 - 2 |c_2|
\end{eqnarray}
in this region. For vanishing magnetic field ferromagnetism is obtained as long
as
\begin{equation}\label{iso_ccond1}
0\leq \frac{c_1}{|c_2|}\leq\frac{1}{2}.
\end{equation}
This is just the isotropic limit of inequality (\ref{finc1}). For $c_1=\frac{1}
{2}|c_2|$ there is a phase transition to a partially ordered state, the Fermi
zone of the 1-strings starts at $\lambda=0$. The Fermi radius increases and 
stays finite for $H\not=0$ moving counter-clockwise towards the vertical 
$c_1$-axis in the $(c_2,c_1)$-plane. We have strictly proven that there is no 
point where it reaches infinity (for $H=0$) unless $c_2=0$. One can see that 
from equation (\ref{tbae2_ii}), because for $\lambda\to\infty$ $h(\lambda)$ 
vanishes much slower than $p(\lambda)$.

Summarizing the facts, (ii) splits into two parts one ferromagnetic 
(\ref{iso_ccond1}) and one with partially ordered ground state whose Fermi 
radius varies from zero to infinity (s. figure 2). So one can say that the 
second coupling $c_2$ works here like an external magnetic field rendering the 
Fermi radius finite as it is for for the homogeneous antiferromagnetic models 
with $0<H<H_{crit}$.

Analytical solutions of equations (\ref{tbae1_ii}) or (\ref{tbae2_ii}) can be 
obtained for large and small Fermi radius. We start with the first and consider
equation (\ref{tbae1_ii}). It is identical to the TBA of $XXX(\frac{1}{2})$ 
model except for the term with $c_2$. We therefore use the technique of paper 
\cite{babu}, see also paper \cite{hamer}, recasting that term (after Fourier
transformation) as a suitable product. We put as usual $y(\lambda)=\epsilon_1
(\lambda+b)$ with $\epsilon_1(b)=y(0)=0$ and use the symmetry of $\epsilon_1
(\lambda)$. After Fourier transformation
\begin{equation}\label{ft}
f(\omega)=\int_{-\infty}^{\infty}e^{i\omega\lambda}f(\lambda)d\lambda
\end{equation}
we write the $c_2$-term on the RHS of equation (\ref{tbae1_ii}) in the form 
$-C(\omega)h(\omega)$ with
\begin{equation}\label{c}
C(\omega) = - 2\pi |c_2| e^{-i\omega b}.
\end{equation}
With the notations of \cite{hamer} the solution in $\omega$-space is given by
\begin{eqnarray}\label{wh_sol}
y_+(\omega) = (1-G_+(\omega))C(\omega) + G_+(\omega)Q_+(\omega).
\end{eqnarray}
This has to be integrated to obtain the radius $b$. We have carried out the 
first integral by deforming it along the cut on the negative imaginary axis
yielding (for $b\gg 1$)
\begin{eqnarray}\label{b_int}
\int_{-\infty}^{\infty}(1-G_+(\omega))C(\omega)d\omega = \frac{2\pi |c_2| 
\sqrt{2}}{b^2}.
\end{eqnarray}
Together with the second part we have the condition 
\begin{eqnarray}\label{b_cond}
H - c_1 2\pi \sqrt{\frac{2\pi}{e}}e^{-\pi b} + \frac{|c_2|4}{b^2} = 0
\end{eqnarray}
which determines $b=b(H,c_1,c_2)$. For its validity we have to ensure 
\begin{equation}\label{val_cond}
H/c_1\ll 1$ \quad \mbox{and} \quad $|c_2|/c_1\ll 1.
\end{equation}
The free energy is calculated via TBA (see e.g. \cite{devega1}):
\begin{eqnarray}\label{free_en}
\fl
\frac{F}{2N} = f_0 - \frac{T}{2}\int_{-\infty}^{\infty}p(\lambda)
\ln \left( 1 + e^{\epsilon_1(\lambda)/T} \right) d\lambda 
- \frac{T}{2}\int_{-\infty}^{\infty}p(\lambda)
\ln \left( 1 + e^{\epsilon_2(\lambda)/T} \right) d\lambda.
\end{eqnarray}
For vanishing temperature the first integral is proportional to $y_+(i\pi)$ 
yielding (in leading order) the first term in (\ref{free_en1}). The calculation
of the second integral is more involved, after one has made use of equation 
(\ref{iso_tbae}). We give only the result for the dominant terms:
\begin{eqnarray}\label{free_en1}
\frac{F}{2N} = f_0'  - \frac{1}{8 c_1 \pi^2}H^2  - \frac{H}{4} - \frac{H}
{4\pi b} + \frac{|c_2|}{\pi^2 b^4}.
\end{eqnarray}
The term proportional to $H^2$ gives $\chi_0=1/(4c_1\pi^2)$ which in some 
sense can be interpreted as half of the value which we found on the conformal 
line. This is the susceptibility for 
\begin{equation}
H \ll |c_2| \ll c_1 \nonumber
\end{equation}
because then $b$ does not depend on the magnetic field. The term $-H/4$ 
describes a constant magnetization for $|c_2|\to0$, which can be found from BAE
directly.

The result for $|c_2|\ll H\ll c_1$ is difficult to interprete, the limit 
$H\to0$ is not allowed in this case.

Unfortunately we did not succeed in calculating $F$ for $T>0$, which meets 
severe difficulties.

We close the consideration of (ii) calculating the free energy and magnetic
susceptibility for small Fermi radius, that is close to the line of transition
to ferromagnetic behaviour. We solve equation (\ref{tbae2_ii}) for 
$H < H_{crit}$ with $H_{crit}$ from above (\ref{Hcrit}).

Making an expansion in powers of the Fermi radius $b$ we see that the integral
on the RHS (except the $\delta$-term) is of power $b^3$. We therefore can 
easily determine 
\begin{equation}
\epsilon_1(\lambda) = \epsilon_1^-(\lambda) \quad \mbox{for} \quad 
|\lambda|\leq b
\end{equation}
up to terms of power $b^2$.

From $\epsilon_1(b)=0$ we have
\begin{equation}\label{b}
b = \sqrt{\frac{H_{crit}-H}{16c_1+2c_2}}
\end{equation}
and
\begin{equation}\label{epsilon}
\epsilon_1(\lambda) = (16c_1+2c_2)(\lambda^2-b^2).
\end{equation}
The free energy per site is given via equation I (3.14)
\begin{eqnarray}\label{free_en2}
\frac{F}{2N} = -\frac{3}{4} H - \frac{2}{\pi} \frac{1}{\sqrt{16c_1+2c_2}}
(H_{crit}-H)^{3/2}
\end{eqnarray}
and hence
\begin{eqnarray}\label{susz2}
\chi = \frac{3}{2\pi} \frac{1}{\sqrt{16c_1+2c_2}} \frac{1}{\sqrt{H_{crit}-H}}.
\end{eqnarray}
This is to be compared with the same value for the usual $XXX(\frac{1}{2})$
Heisenberg model
\begin{eqnarray}
\chi_{XXX} = \frac{2}{\pi} \frac{1}{\sqrt{16c}} \frac{1}{\sqrt{H_{crit}-H}}
\end{eqnarray}
in our normalization of coupling constant. That result is, of course, not the 
limit $c_2\to0$ of equation (\ref{susz2}), because the change in I (3.14) also
must be taken into account.

At the end of this section we shortly comment on the sector (iii). It is 
treated in the same way as above the sector (ii). Equations (\ref{tbae1_ii}) 
and (\ref{tbae2_ii}) are replaced by 
\begin{eqnarray}\label{tbae1_iii}
\fl
\epsilon_2(\lambda) = - 2\pi c_2 p(\lambda) + \frac{\pi|c_1|\lambda}
{\sinh(\pi\lambda)} + \left( \frac{\lambda}{2\sinh(\pi\lambda)} + h(\lambda) 
\right)*\epsilon_2^- + \frac{H}{2},
\end{eqnarray}
\begin{eqnarray}\label{tbae2_iii}
\fl
\epsilon_2^+(\lambda) = |c_1|a_2(\lambda) - c_2(a_1(\lambda)+a_2(\lambda)) -
\left( \delta(\lambda) + \frac{2a_2(\lambda)+a_4(\lambda)}{2\pi}\right)*
\epsilon_2^- + 2H. 
\end{eqnarray}
The critical magnetic field can be read off from equation (\ref{tbae2_ii}):
\begin{equation}\label{Hcrit1}
H_{crit} = \frac{8}{3} c_2 - |c_1|.
\end{equation}
For vanishing field we have ferromagnetic behaviour as long as
\begin{equation}\label{iso_ccond2}
\frac{|c_1|}{c_2}\geq\frac{8}{3}
\end{equation}
(compare equation (\ref{finc4})).

The power expansion in $b$ is rather simple while the Wiener-Hopf calculation
for large $b$ is a little bit more involved. Therefore it is not carried out 
here. The results will qualitatively agree with those from above.

The phase structure is depicted in figure 2. We have four sectors and two 
singular lines (the positive axes). Three of the sectors show critical 
behaviour without mass gap; one of them is truly antiferromagnetic with 
infinite Fermi zone. The remaining one is ferromagnetic.

\begin{figure}
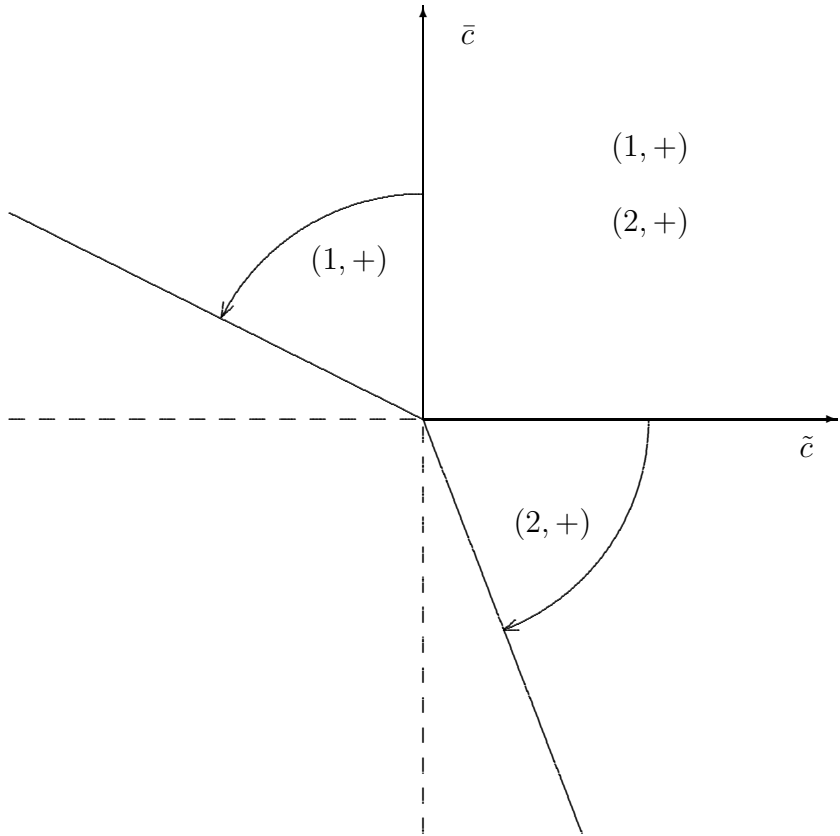

\input figure2.tex
\caption{The phase structure of the $XXX(\frac{1}{2},1)$ model. The 
sector without indication of ground state strings is ferromagnetic. Again axes 
are drawn broken except if they coincide with sector borders, where they are 
drawn solid.} 
\end{figure}

\section{Conclusions}
We have considered the $XXZ(\frac{1}{2},1)$ model with strictly alternating
spins in one part of its critical region of anisotropy $0\leq\gamma<\pi/2$.
We expect a similar but not identical behaviour in the other part $\pi/2<
\gamma<\pi$ because there is no obvious symmetry between the two regions. The 
model contains two parameters, anisotropy and the ratio of coupling constants,
and shows a rich physical structure. Except for the isotropic model (where we 
have a ferromagnetic region) we found an antiferromagnetic ground state and no
mass gap. So the model behaves critically, but it is conformally invariant only
on a line $\bar{c}=\tilde{c}$ and also in a large sector including this line 
and having at least one negative coupling. Around that line there exist sectors
where the ground state does not depend on $\gamma$, seperated from each other 
by sectors where it depends crucially on $\gamma$. It is remarkable that two
kinds of sectors are also different with respect to the occurence of finite 
Fermi zones.

The sectors around the line with equal couplings are well studied now, their 
ground states and excitations have been established. At the line $\bar{c}=
\tilde{c}$ we have calculated low temperature heat capacity and magnetic 
susceptibility. Different signs of couplings cause very different behaviour,
e. g. different central charges (1 or 2) and different behaviour of 
susceptibilities.

A subsequent paper will deal with finite size corrections in those sectors.
We expect the standard results in the conformal case while apart from that 
conformal symmetry does not make any prediction.

The sectors with finite Fermi zones require further treatment, including 
numerical studies. The same applies to heat capacity and susceptibility for
different coupling constants.

\section*{Acknowledgement}
We thank H. M. Babujian for helpful discussions.

\newpage
\section*{Figure captions}
{\bf Figure 1.} The phase structure of the $XXZ(\frac{1}{2},1)$ model for 
$\gamma=\pi/3$. The ground state strings are indicated for the four different
sectores. The arrows symbolize the decreasing Fermi radii of the corresponding
strings. Broken lines are coordinate axes. Where axes are drawn solid they 
coincide with sector borders. \\[2cm]
{\bf Figure 2.} The phase structure of the $XXX(\frac{1}{2},1)$ model. The 
sector without indication of ground state strings is ferromagnetic. Again axes 
are drawn broken except if they coincide with sector borders, where they are 
drawn solid.

\end{document}